\documentclass[journal,10pt,two column]{IEEEtran}
\IEEEoverridecommandlockouts
\usepackage{cite}
\usepackage{amsmath,amssymb,amsfonts}
\usepackage{algorithmic}
\usepackage{graphicx}
\usepackage{textcomp}
\usepackage{xcolor}
\usepackage{subfigure}
\usepackage{fancyhdr}
\usepackage{marvosym}
\def\BibTeX{{\rm B\kern-.05em{\sc i\kern-.025em b}\kern-.08em
    T\kern-.1667em\lower.7ex\hbox{E}\kern-.125emX}}
\begin{document}
\title{Design of Full-Duplex Millimeter-Wave Integrated Access and Backhaul Networks\\}
\author{Junkai~Zhang,~\IEEEmembership{Student~Member,~IEEE}, Navneet~Garg,~\IEEEmembership{Member,~IEEE,} Mark~Holm,~\IEEEmembership{Member,~IEEE,}
and~Tharmalingam~Ratnarajah,~\IEEEmembership{Senior~Member,~IEEE}
\thanks{This work has been accepted for publication in the special issue of Full Duplex Communications Theory, Standardization and Practice, IEEE Wireless Communications, February, 2021. Manuscript received May 13, 2020; revised September 3, 2020; accepted October 19, 2020. Date of current version February 12, 2021.}%
\thanks{J. Zhang, N. Garg, and T. Ratnarajah are with Institute for Digital Communications, The University of Edinburgh, Edinburgh, EH9 3FG, UK (e-mail: \{jzhang15, ngarg, T. Ratnarajah\}@ed.ac.uk.}%
\thanks{M. Holm is with Radio Basestation Systems Department, Huawei Technologies (Sweden) AB, Gothenburg, Sweden. (e-mail: mark.holm@huawei.com).}
}

\maketitle
\pagestyle{empty}

\section*{Abstract}
One of the key technologies for the future cellular networks is full duplex (FD)-enabled integrated access and backhaul (IAB) networks operating in the millimeter-wave (mmWave) frequencies. The main challenge in realizing FD-IAB networks is mitigating the impact of self-interference (SI) in the wideband mmWave frequencies. In this article, we first introduce the 3GPP IAB network architectures and wideband mmWave channel models. By utilizing the subarray-based hybrid precoding scheme at the FD-IAB node, multiuser interference is mitigated using zero-forcing at the transmitter, whereas the residual SI after successfully deploying antenna and analog cancellation is canceled by a minimum mean square error baseband combiner at the receiver. The spectral efficiency (SE) is evaluated for the RF insertion loss (RFIL) with different kinds of phase shifters and channel uncertainty. Simulation results show that, in the presence of the RFIL, the almost double SE, which is close to that obtained from fully connected hybrid precoding, can be achieved as compared to half duplex systems when the uncertainties are of low strength.

\section*{Introduction}
\label{introduction}
\noindent Key technologies, namely, millimeter-wave (mmWave) wideband communications, full duplex (FD) transmissions, and integrated access and backhaul (IAB) networks, are emerging as the backbone of 5G and beyond communications. The large bandwidth provided by mmWave systems can be exploited for wideband transmissions to increase data rates, which are orders of magnitude more than that of the current microwave systems. However, a beamformed array with a large number of antennas is needed to compensate for the higher path loss at mmWave frequencies \cite{7448873}. Moreover, to enhance the coverage, dense deployment of multi-antenna access points has been considered as a promising approach. However, providing traditional fiber backhauling connection to all these small cells is not possible either economically or physically. To address this issue, the 3rd Generation Partnership Project (3GPP) proposed cost-effective dense deployment of wireless backhauling through IAB nodes to achieve promising gains even under higher mobile data traffic \cite{DBLP:journals/corr/abs-1906-09298}.

Moreover, to leverage the full benefits of IAB networks with the mmWave wideband, the IAB nodes are set to operate in the FD mode. Compared to half duplex (HD) transmission, FD can enhance the spectral efficiency (SE) and reduce the communication delay without any requirement for the guard time/band \cite{8246856}. Unlike traditional microwave communications, where full digital baseband (BB) precoding schemes are sufficient, hybrid precoding is essential in mmWave communications \cite{7448873}. For wideband mmWave-FD-IAB networks, hardware-efficient subarray-based hybrid precoding is adopted in this article.

Since in an FD-IAB network the access and backhaul communications occur at the same time-frequency resource, it naturally gives rise to self-interference (SI) at the receiver of the FD-IAB node. Typically, the magnitude of the SI can be more than 100 dB stronger than the signal of interest, as studied in \cite{7414127}. Such a high SI power can significantly exceed the hardware dynamic range and distort the benefits of FD transmission. Thus, it is important to reduce SI power before down-conversion. In microwave communications, successful SI cancellation (SIC) can be achieved at the antenna domain (i.e., by deploying special antenna isolation), the RF domain (i.e., by replicating the SI channel and subtracting it from the received signal), and the digital domain (i.e., by canceling the residual SI [RSI] after RF cancellation by beamformer design). Usually, a combination of these stages has shown satisfactory results \cite{8246856}, which we also expect to provide a good solution for mmWave wideband communications. In this article, we mainly focus on the design of digital cancellation, where antenna isolation and RF cancellation are assumed to be successfully achieved. Therefore, only the RSI signal will be handled in the digital domain.

In this article, we first introduce the fundamental 3GPP network architectures for FD-IAB systems, followed by a description of the general mmWave and SI channel models. Next, a hybrid analog/digital transceiver design via the cost-efficient subarray structure for the multiuser scenario is explained. The multiuser interference (MUI) at the transmitter of the IAB node and the RSI at the receiver of the IAB node are mitigated by zero-forcing (ZF) and minimum mean squared error (MMSE) in the digital BB domain, respectively. Further, the performance limitations of FD-enabled multiuser mmWave-IAB networks under subarray hybrid precoding structure are studied in the presence of the RF insertion loss (RFIL) and the channel estimation error (CEE). With the RFIL, simulations show that the SE performance of the fully connected hybrid precoding structure is similar to that for the subarray-based hybrid precoding structure. Moreover, as the CEE increases, the rate improvement of FD over HD decreases. Besides, the SE intersection point of FD and HD that appears at the backhaul link enables the understanding of the maximum achievable digital cancellation, which will encourage the development of advanced hybrid transceivers with efficient resource allocation schemes in the future.
\begin{figure}[t!]
\centering
\subfigure[]{
\includegraphics[width=\columnwidth]{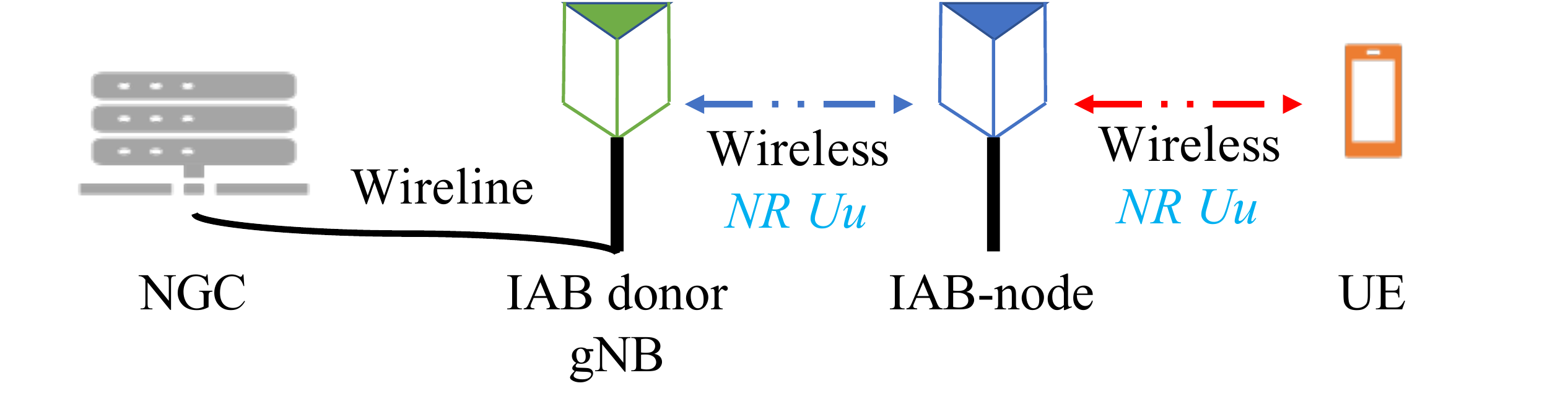}\label{a}}
\subfigure[]{
\includegraphics[width=\columnwidth]{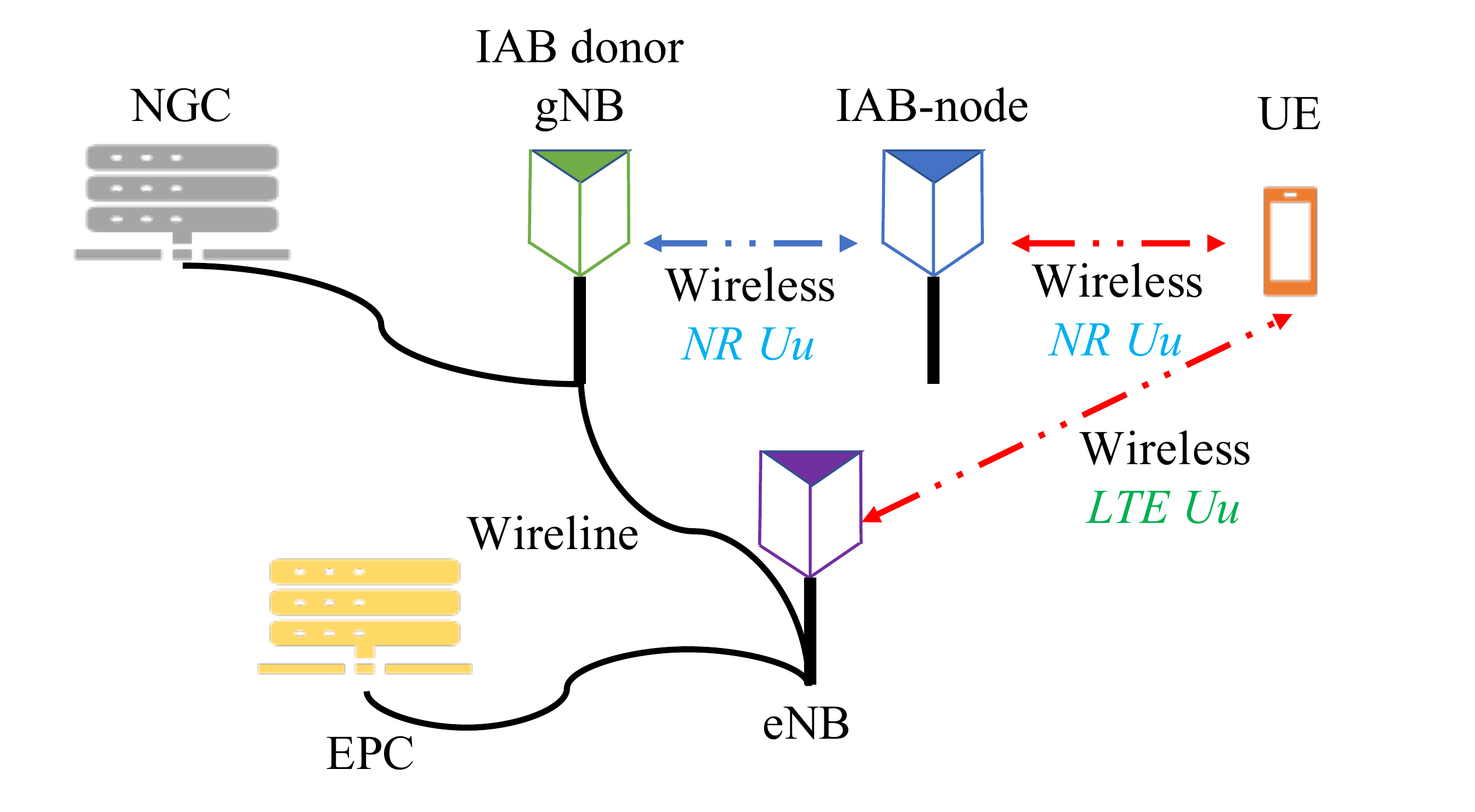}\label{b}}
\subfigure[]{
\includegraphics[width=\columnwidth]{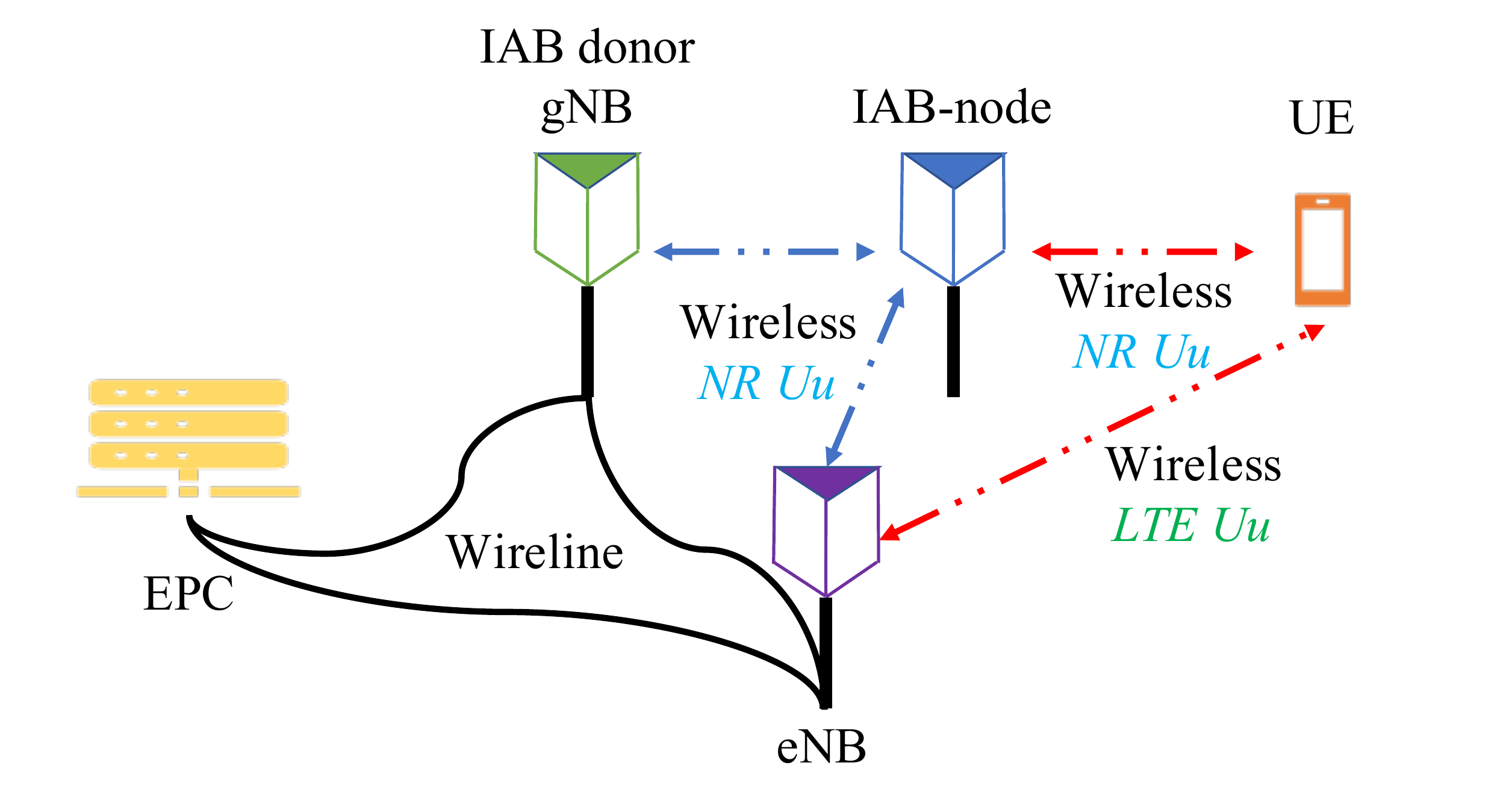}\label{c}}
\caption{Examples for FD-IAB network architectures operating in SA mode and NSA mode: a) UE: SA with NGC, IAB node: SA with NGC; b) UE: NSA with EPC, IAB node: SA with NGC; c) UE: NSA with EPC, IAB node: NSA with EPC.}
\label{3GPP1}
\end{figure}

\section*{3GPP Network Architectures}
3GPP Release 16 explores the standards for 5G New Radio (NR) communications. IAB architectures, radio protocols, and physical layer aspects related to relaying of access traffic by sharing radio resources between access and backhaul links are investigated in the technical specification TR 38.874 \cite{3gpp}. These initial studies show the benefits of in-band backhauling over out-of-band backhauling for access links. However, these fundamental results for FD operations are still in their infancy. Further, knowledge of the impact of FD operations at mmWave frequencies is also limited, since the wideband channel model for FD operations still needs thorough investigation. According to the 3GPP specification in \cite{3gpp}, IAB systems are typically deployed in two modes, namely standalone (SA) mode and non-standalone (NSA) mode, as shown in Fig.~\ref{3GPP1}. In the SA mode shown in Fig.~\ref{a}, the IAB node connects to the 5G next-generation core (NGC) network via the IAB donor (gNB), and the user equipment (UE) also operates in the SA mode (i.e., it only connects to the IAB node). In Fig.~\ref{b}, the UE is connected in the NSA manner, while the IAB node is in the SA mode. In this scenario, both Long Term Evolution (LTE) radio and NR can be used for the UE, and NR links are utilized for backhauling. Further, if the IAB node works in the NSA mode, it is also connected to the eNB nodes (i.e., the 4G base stations), as shown in Fig.~\ref{c}. Thus, a UE in the NSA mode can choose to connect the IAB-connected eNB or a different one. In the third scenario, the IAB node can utilize the LTE links for initial access, route selection, and so on.

\begin{figure}
\centering
\subfigure[]{
\includegraphics[width=\columnwidth]{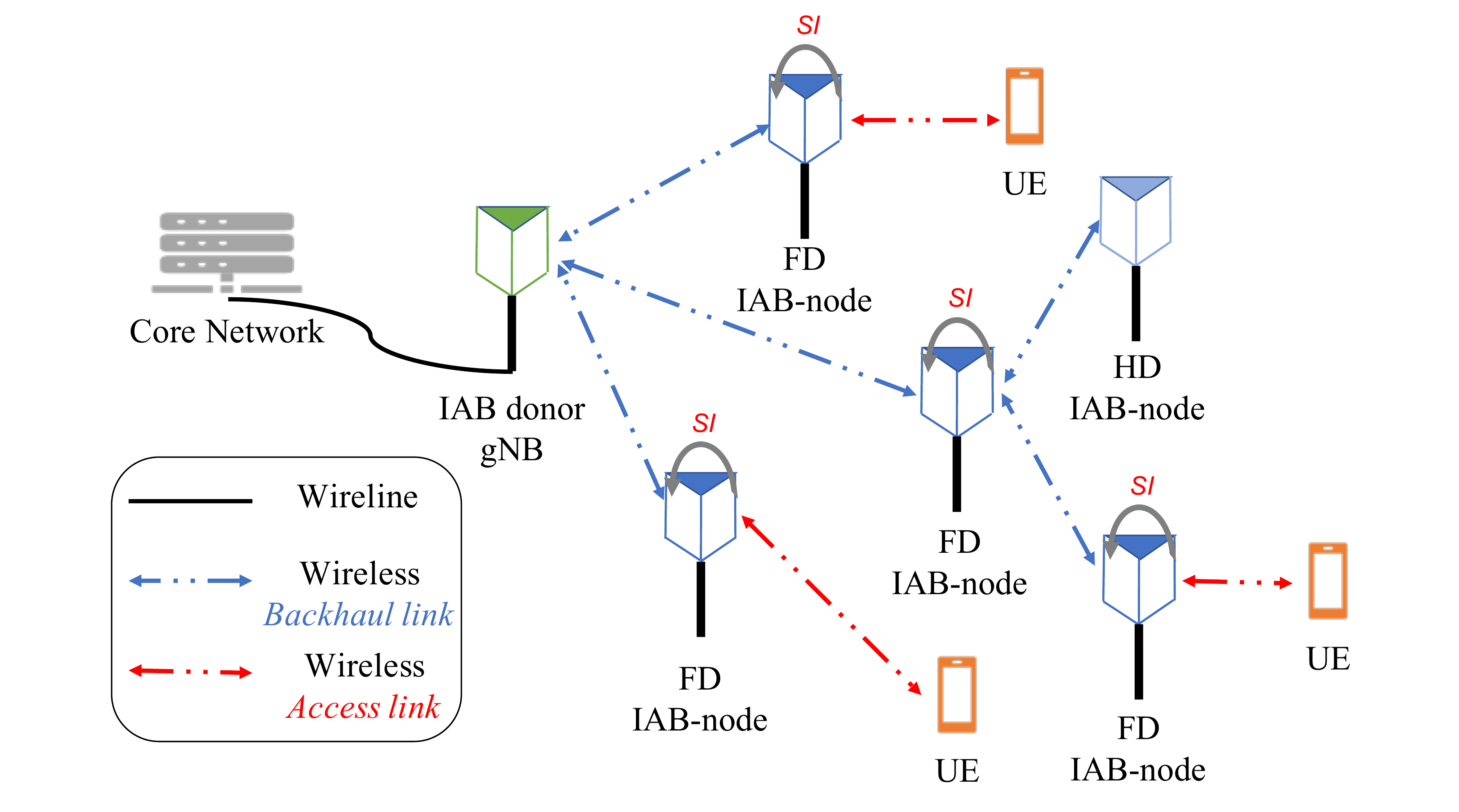}\label{3GPP2}}
\subfigure[]{
\includegraphics[width=\columnwidth]{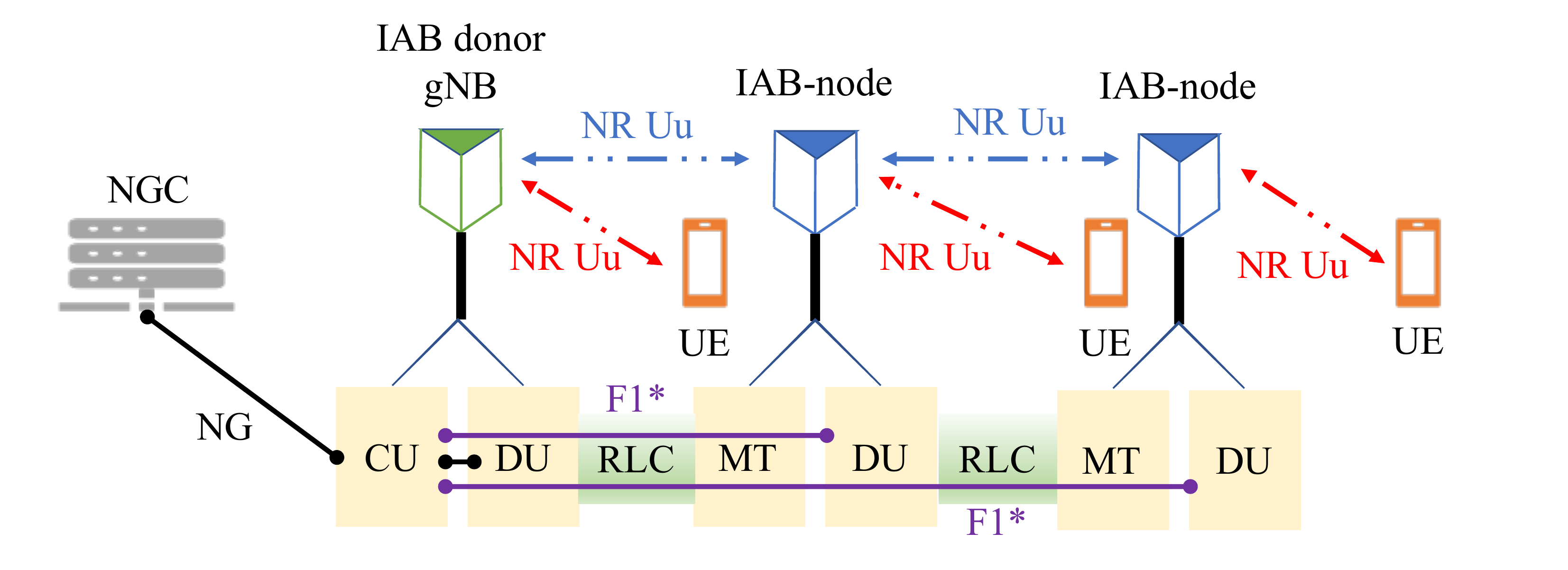}\label{arc}}
\caption{a) Illustration of multihop mmWave-FD-IAB network architecture diagram in SA mode; b) CU/DU split architecture for multihop IAB system.}
\end{figure}

A multihop mmWave IAB networks in SA mode is shown in Fig.~\ref{3GPP2}. In this figure, there are three kinds of nodes listed as follows,
\begin{itemize}
    \item A single logical IAB donor, which is the source node, also known as the gNB. It takes responsibility for functionality and splits according to the 3GPP next generation radio access network (NG-RAN) architecture \cite{3gp}. Usually, the gNB has a wired connection to the core network (NGC) and has wireless connections to other nodes.
    \item IAB nodes, which wirelessly communicate with both backhaul and access links, provide FD operations and perform IAB-specific tasks such as resource allocation, route selection, and optimization. The IAB nodes can be connected to other HD-IAB nodes or FD-IAB nodes.
    \item UE nodes, which request and receive the con- tents via FD or HD operation. Since UEs operate in the SA mode, they only connect to the IAB nodes.
\end{itemize}
Typically, the IAB node enables not only UEs but also other FD/HD-IAB nodes to communicate with the gNB. In the SA architecture illustrated in Fig~\ref{3GPP2}, IAB nodes forward their own backhaul traffic to the core network in different spectrum, whereas with this general star topology, Taghizadeh \textit{et al}. \cite{8638843} consider a central station delivering the backhaul traffic from multiple nodes, which may require efficient interference management schemes.

There are two kinds of topology models to characterize such multihop networks. The first one is the spanning tree (ST) model, where one IAB node connects to only one parent node (i.e., the IAB donor or another IAB node). The second model uses directed acyclic graphs (DAGs), where one IAB node has multiple parent nodes, or has multiple routes to one parent node, or a combination of these two cases \cite{DBLP:journals/corr/abs-1906-01099,3gpp}. For the multihop IAB networks, a simple and low-complexity architecture, the central unit (CU)/distributed unit (DU) split architecture, is preferred in studies \cite{DBLP:journals/corr/abs-1906-01099,DBLP:journals/corr/abs-1906-09298}, and is shown in Fig.~\ref{arc}, where the CU and DU represent external interfaces of the node. In this architecture, the IAB node has two NR functional units: the mobile termination (MT) unit, which controls the upstream link connection with the IAB donor or the IAB node; and the DU, which provides connections to UEs or MTs on other IAB nodes of the downstream link. The IAB donor has two functional units as well: the CU is responsible for serving the DUs on all IAB nodes and the donor itself, while the DU provides support to the UEs and MTs of downstream IAB nodes. The F1* function connects the interface of the IAB node to the interface of the IAB donor. It runs on the radio link control (RLC) channels, rep- resenting the connections between the DU and the downlink MT or UEs.

\section*{Channel Models}
\subsection*{General mmWave Channel}
The mmWave channel has several characteristics that differentiate it from the traditional microwave channels, such as higher path loss (due to higher operating frequencies), the spatial selectivity (due to high path losses and beamforming), and increased correlation among antennas (due to densely collocated arrays). These distinctive characteristics imply that the statistical fading distributions such as the Rayleigh distribution used in traditional wireless channels become inaccurate, since the number of fading paths is small. Hence, the mmWave channel between two different nodes is likely modeled as a geometric wideband frequency-selective channel according to the extended Saleh-Valenzuela model, studied in \cite{7448873,744887}.

An orthogonal frequency-division multiplexing (OFDM) system with $K$ subcarriers is adopted, where $D$ cyclic prefix (CP) is added to avoid the inter-symbol interference (ISI). For each of the $D$ taps of the wideband channel, scatterers in the area contribute to multiple propagation paths. These reflected multipath components (rays) arrive in clusters, which cause the sparse nature of the channel response. The value of the channel at tap $d = 1, 2,\ldots, D$ is modeled using the product of the complex random gain, the complex exponential of angles of arrival and departure (AoAs/AoDs), and the pulse-shaping filter. The complex random gain of each ray has the magnitude following the Rayleigh distribution with the parameter defined by the number of total paths. For the uniform planar arrays (UPAs), the central azimuth AoAs/AoDs of fading paths (rays) in each cluster are uniformly distributed in $[-\pi,\pi]$, and the corresponding central elevation AoAs/AoDs are uniformly distributed in $ [-\pi/2,\pi/2]$. In each cluster, these azimuth and elevation angles of the rays are assumed to have Laplacian distribution with a given angle spread. The raised cosine pulse shaping filter is utilized with sampling time $T_s$, evaluated at $dT_s-\tau_{c,l}$ seconds, where $\tau_{c,l}$ is the path delay of the $l$th ray in the $c$th cluster and is uniformly distributed in $[0,DT_s]$. The close-in (CI) path loss model with a reference distance of 1m is introduced to capture the average path loss. Ultimately, the channel at subcarrier $k=1, 2, ..., K$ is given by the discrete Fourier transform (DFT) of the delay-$d$ channel.

\subsection*{Self-Interference Channel}
The FD-IAB node comprises a transmit antenna array and a receive antenna array. In FD operations, an mmWave SI channel is defined as the mmWave channel between the transmit antenna and the receiver antenna at the IAB node. Through measurements, the mmWave SI channel is verified to have both line-of-sight (LoS) and non-line-of-sight (NLoS) components \cite{7414127}. The LoS component accounts for deterministic direct path loss. Its strength is very high due to a very short distance between the transceiver of the IAB node and is assumed to adopt a near-field model, since the distance between the transceivers is smaller than $2D^2/\lambda$, where $D$ is the antenna aperture diameter, and $\lambda$ is the wavelength \cite{8246856}. The coefficient of the LoS channel matrix depends on the distance between the individual elements of the transceiver. The NLoS component indicates random components caused by reflections from obstacles around the IAB node, where the general mmWave channel model may be accept- able, except with a smaller number of rays. A Rician-alike channel model could be utilized to model the SI channel due to a strong LoS path. A detailed hypothetical wideband mmWave SI channel model is formulated in our recent work  \cite{Zhan2012}. It is worth noting that there is still ambiguity in characterizing the mmWave SI channel model in the literature.

A study in \cite{8246856} shows that the resulting SI channel is sparse and low rank. Unfortunately, as mentioned in \cite{article}, the difficulties of SIC arise due to its inability to cancel the NLoS component of the SI signal by the three-stage SIC scheme. It is due to the fact that the present SI channel estimation methods have proved to be inaccurate due to the strong antenna correlation in the near-field region. Moreover, in general, the channel estimation for microwave communications assumes steady oscillator phase noise (PN); however, for mmWave communications, this assumption can cause large estimation error, since the PN changes rapidly and cannot be ignored. In \cite{article}, with the Rician SI channel model, a joint SI channel and PN estimation algorithm for mmWave communications using the Kalman filter is proposed, which is shown to achieve its mean squared error (MSE) lower bound successfully. With an efficient estimator, the RSI can be decreased to an accept- able amount.

However, it is difficult to estimate the large and
sparse mmWave MIMO channel in reality. Therefore, the CEE, $\mathbf{\Delta}_{\mathrm{SI}}[k]$, is introduced to model the imperfect RF effective SI channel and analyze the corresponding system performance. The perfect RF effective SI channel (i.e., the product of the RF combiner, the SI channel matrix, and the RF precoder) at the $k$th subcarrier is assumed to be the sum of the estimated RF effective SI channel and the random CEE. The CEE is assumed to be Gaussian with zero mean and small variance \cite{5595019}. However, interference leakage occurs due to the CEE and results in the RSI power. The impact of the CEE on the system capacity is given in a later section.

\begin{figure*}[t!]
\centering
\subfigure[]{
\includegraphics[width=2\columnwidth]{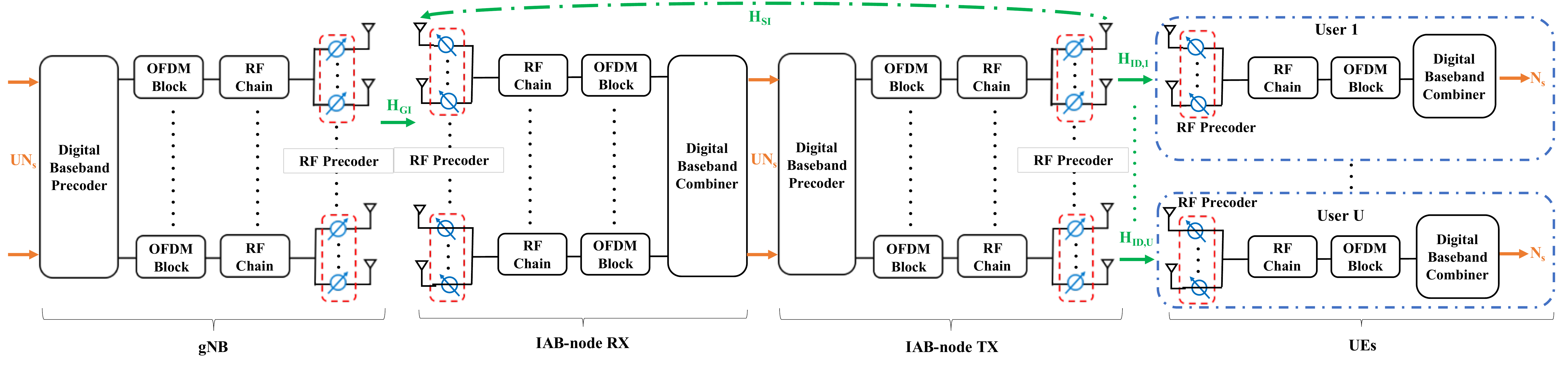}\label{hp}}
\end{figure*}
\begin{figure}
\centering
\subfigure[]{
\includegraphics[width=0.9\columnwidth]{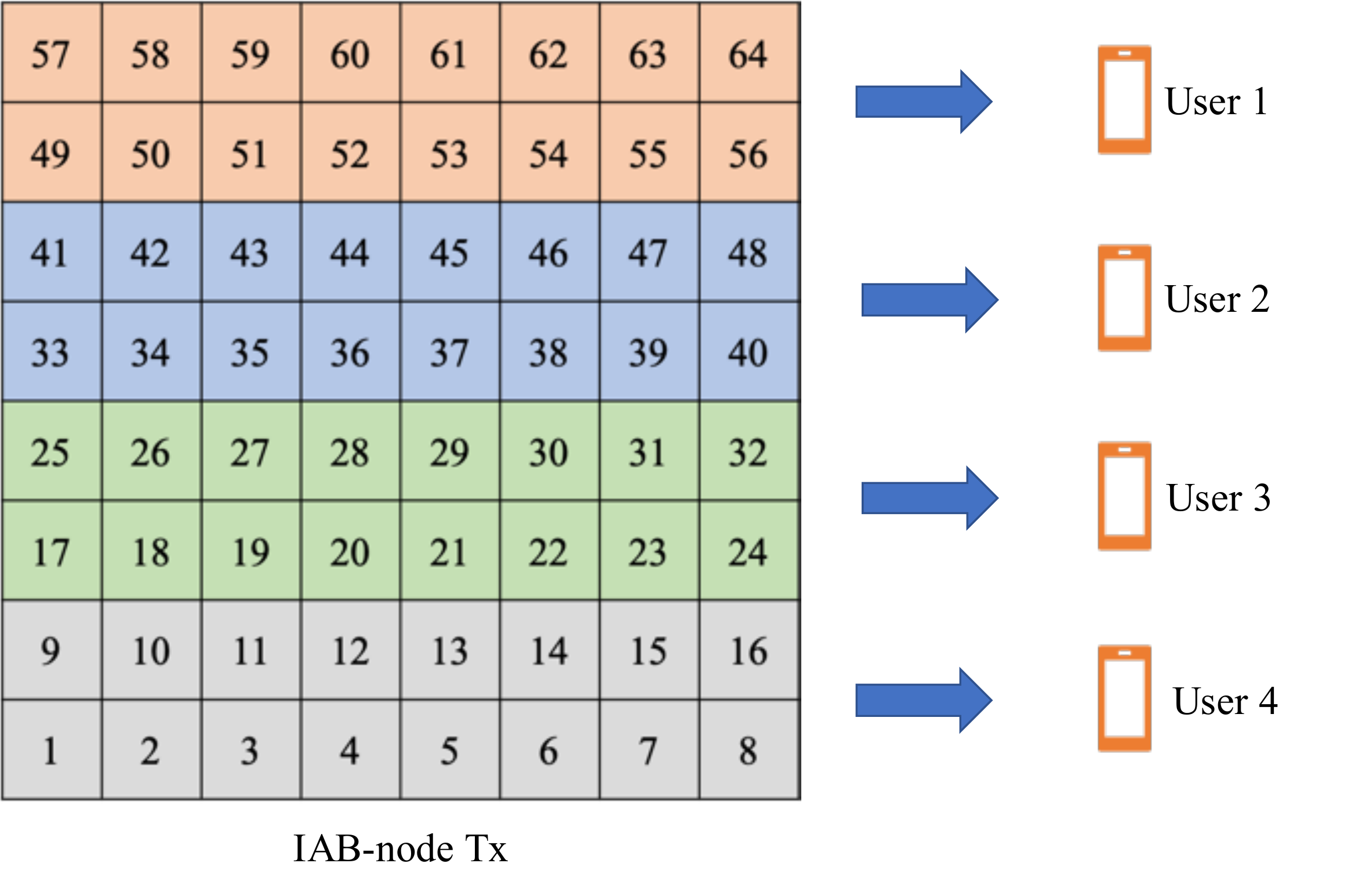}\label{sub}}
\caption{a) Multiuser hybrid transceiver for FD-IAB wideband mmWave system; b) subarray structure for multiuser transmission (8$\times$8 UPA, 4 users).}\label{noo}
\end{figure}
\section*{Hybrid Transceiver Design}
Since the wideband channel is frequency-selective, each node adopts an OFDM system, ensuring that each subcarrier experiences a flat-fading channel. In conventional MIMO networks, only BB beamforming has been used to maximize the SE, provided that each node has a fully connected RF chain corresponding to each antenna. However, in mmWave communications, the small aperture size of the antenna and the large array size make it impossible for each antenna to have an RF chain. Thus, hybrid precoding has been utilized with a much lower number of RF chains than the number of antennas (e.g., for gNB with 256 antennas, the number of RF chains is set to 4). For the wideband channel, we assume the BB beamforming is different for each subcarrier and is based on the number of RF chains and that of data streams. In contrast, the RF beamforming is achieved via phase shifters (PSs) and is the same for all subcarriers. The dimension of the RF beam- forming is defined by the number of RF chains and that of antenna arrays. There are two kinds of hybrid transceiver structures studied in \cite{7448873},
\begin{itemize}
    \item Fully connected, where each RF chain connects to each antenna(i.e., all the antennas are connected to each of the RF chains)
    \item Partially connected (or subarray), where each RF chain only connects to a disjoint subset of antennas
\end{itemize}
Although both structures employ fewer RF chains, the second structure is easier to deploy and more cost-efficient in practice. This is because in a fully connected structure, mmWave antenna spacing and aperture size are small, which causes a high correlation between the outputs of RF chains. For the multiuser scenario, each subarray is set to serve a single user, which means that the number of subarrays can be selected based on the number of users (Fig.~\ref{noo}). In Fig.~\ref{sub}, each user is shown to be served by 1 subarray with a 16-antenna-array panel.

Figure~\ref{hp} gives the architecture of a multiuser hybrid transceiver for an FD-IAB wideband mmWave system, which is used for the analysis for the IAB networks in this work. For the transmitter side, the OFDM block performs inverse discrete Fourier transform (IDFT) and adds the CP to the precoded streams using the BB precoder. On the receiver side, the OFDM block removes the CP and performs the DFT, followed by the BB combiner operation. Since each of the users (say $U$ in total) communicates single data streams, the total number of data streams should not exceed the number of RF chains at the transmitter of the IAB node.

Our objective in hybrid transceiver design is to maximize the SE across all subcarriers for access and backhaul links. This joint maximization problem concerning the RF and BB precoders and combiners has a few constraints, as follows. Since RF precoders and combiners are implemented using PSs, it poses the constraint that the magnitude of each entry of the RF precoder and combiner matrices should be precisely equal to 1. Further, the effective coupled RF and BB precoders must satisfy the transmit power constraint. Assuming equal power allocation across all data streams, the squared norm of the hybrid precoder at each subcarrier should not exceed the length of the data stream vector. Since the maximization problem is non-convex due to coupled RF and BB variables, a joint optimal solution for these variables is intractable.

Interestingly, the near-optimal solution, where the RF and the BB variables are obtained separately, is studied in \cite{7448873}. Ideally, the RF part of the hybrid precoders or combiners is computed as the dominant eigenvector corresponding to the eigenvalue decomposition (EVD) of the channel correlation matrix (i.e., the sample covariance matrix). For the RF precoder at the transmitter, the sample transmit covariance matrix is computed, while for the RF combiner at the receiver, the sample receive covariance matrix is used. In addition to this, the easier implementation of the subarray structure simplifies the precoder and combiner design to a block diagonal form, which incurs lower computational complexity. Thus, for the subarray-based structure, RF variables are obtained using the correlation matrix of the sub-channel matrix corresponding to the antenna elements of the subarray. Note that EVD incurs a cubic computational overhead (say $O(N^3)$). Thus, the subarray structure reduces the overhead to $O((N/U)^3)$. However, the optimal solution above needs to access the channel state information (CSI) of the large mmWave channel, which is difficult to estimate in reality. Therefore, in this article, for both the backhaul and access links, we assume the accurate knowledge of RF effective channel only, where the optimal RF precoders and combiners are provided by genie. In practice, these RF quantities are obtained by beam-training codebooks \cite{Zhan2012}. The optimal BB precoders/combiners can then be obtained as the left/right dominant singular vectors of the effective channel matrix. Note that the above BB transceiver design is applicable for the nodes, which have perfect interference cancellation or operate in HD mode. However, in FD-IAB networks, there is strong SI present at the IAB node that needs cancellation. Thus, the above hybrid design for the IAB node needs to be modified.
 
\subsection*{Multiuser Interference and Self-Interference Cancellation}
To maximize the SE of the multiuser FD-IAB network, the BB precoders/combiners at the IAB node must achieve the following. The transceiver design should:
\begin{itemize}
    \item Mitigate the RSI at the receiver of the IAB node
    \item Cancel the MUI at the transmitter of the IAB node
\end{itemize}

Technically, in mmWave, such a high-power SI is likely to exceed the limitation of the dynamic range on analog-to-digital converters (ADCs) and results in a stronger nonlinear signal than that of the desired signal. Therefore, the antenna and RF cancellation are adopted before the digital process to cancel out a large amount of SI \cite{7224732}. However, the study in \cite{8246856} states that for mmWave wideband, RF cancellation faces difficulties in the canceller design due to the realization of a large number of taps and the high delay spread of the SI channel, and also experiences severe performance degradation due to RF impairments as compared to that in microwave communications. Wideband active analog SIC is studied in \cite{luo}. With this novel RF cancellation technique, those difficulties of traditional RF canceller design can be overcome.

Consequently, the remaining RSI will be handled by digital cancellation, that is, by applying the MMSE BB combiner at the IAB node. In order to achieve good digital SIC, the number of RF chains at the receiver of the IAB node should be at least the sum of the number of data streams transmitted and received by the IAB node. Since the BB SIC depends on the estimated channel state information (CSI) of the RF effective SI channel, the CEE has a strong impact on the performance of digital SIC. Staged SIC that combines the RF and digital cancellation is studied in our recent work \cite{Zhan2012}. Regarding the MUI, traditional ZF is utilized at the IAB node transmitter to obtain the desired BB precoder.

\subsection*{RF Insertion Loss}
The RFIL, $L_{\mathrm{RF}}$,  which is caused by PSs, power dividers (PDs), and power combiners (PCs), is an important loss that cannot be easily compensated by the existing technologies in mmWave. Failure to take the RFIL into account may result in higher analytical spectral efficiency. To assess the impact of the RFIL, the factor, $1/\sqrt{L_{\mathrm{RF}}}$, is multiplied with the RF precoder/combiner matrices.

For the fully connected structure, the RF precoding requires $N_{\mathrm{RF}}$ PDs ($N_\mathrm{t}$-way), $N_\mathrm{t}$ PCs ($N_{\mathrm{RF}}$-way) and $N_\mathrm{t}N_{\mathrm{RF}}$ PSs, while the RF combining needs $N_\mathrm{r}$ PDs ($N_{\mathrm{RF}}$-way), $N_{\mathrm{RF}}$ PCs ($N_\mathrm{r}$-way), and $N_\mathrm{r}N_{\mathrm{RF}}$ PSs, where $N_\mathrm{t}$, $N_\mathrm{r}$ and $N_{\mathrm{RF}}$ denotes the number of transmitters, receivers, and RF chains, respectively.

On the other hand, for the RF precoding with $U$ subarrays, $U$ PDs ($N_\mathrm{t}/U$-way) and $N_\mathrm{t}$ PSs are required, while at each subarray (user) of the receiver, a PC ($N_\mathrm{r}/U$-way) and $N_\mathrm{r}/U$ PSs are needed. Specially, at the receiver of the IAB-node, $N_\mathrm{r}$ PDs ($N_{\mathrm{RF}}/U$-way), $N_{\mathrm{RF}}N_\mathrm{r}/U$ PSs, and $N_{\mathrm{RF}}$ PCs ($N_\mathrm{r}/U$-way) are required.

Given that a cascade of $\lceil \mathrm{log}_2(X)\rceil$ stages of 2-way PDs and $\lceil \mathrm{log}_2(Y)\rceil$ stages of 2-way PCs are utilized to construct the $X$-way PD and the $Y$-way PC, respectively. $L_{\mathrm{RF}}$ is given by the product of the static power loss of PDs (i.e., $P_\mathrm{D}\lceil \mathrm{log}_2(X)\rceil$ dB), PSs (i.e., $P_\mathrm{PS}$ dB), and PCs (i.e., $P_\mathrm{C}\lceil \mathrm{log}_2(Y)\rceil$ dB), where $P_\mathrm{D}=0.6$ dB and $P_\mathrm{C}=3.6$ dB denote the power loss of the PD and the PC, respectively. Moreover, there are two kinds of PSs, i.e., the active PS ($P_\mathrm{PS}=-2.3$ dB) and the passive PS ($P_\mathrm{PS}=8.8$ dB) \cite{8333733}.

\begin{figure}
\centering
\subfigure[]{
\includegraphics[width=\columnwidth]{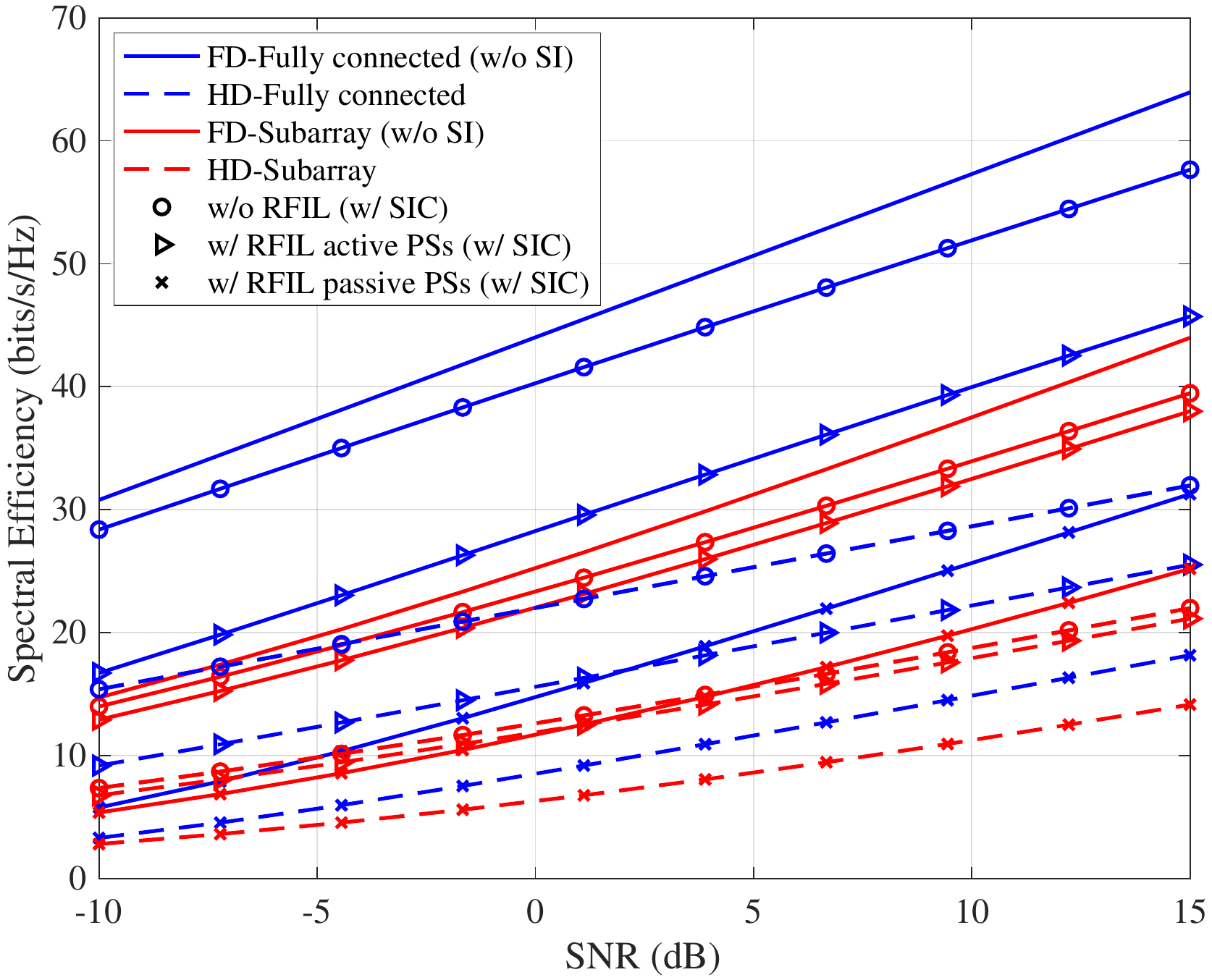}}
\subfigure[]{
\includegraphics[width=\columnwidth]{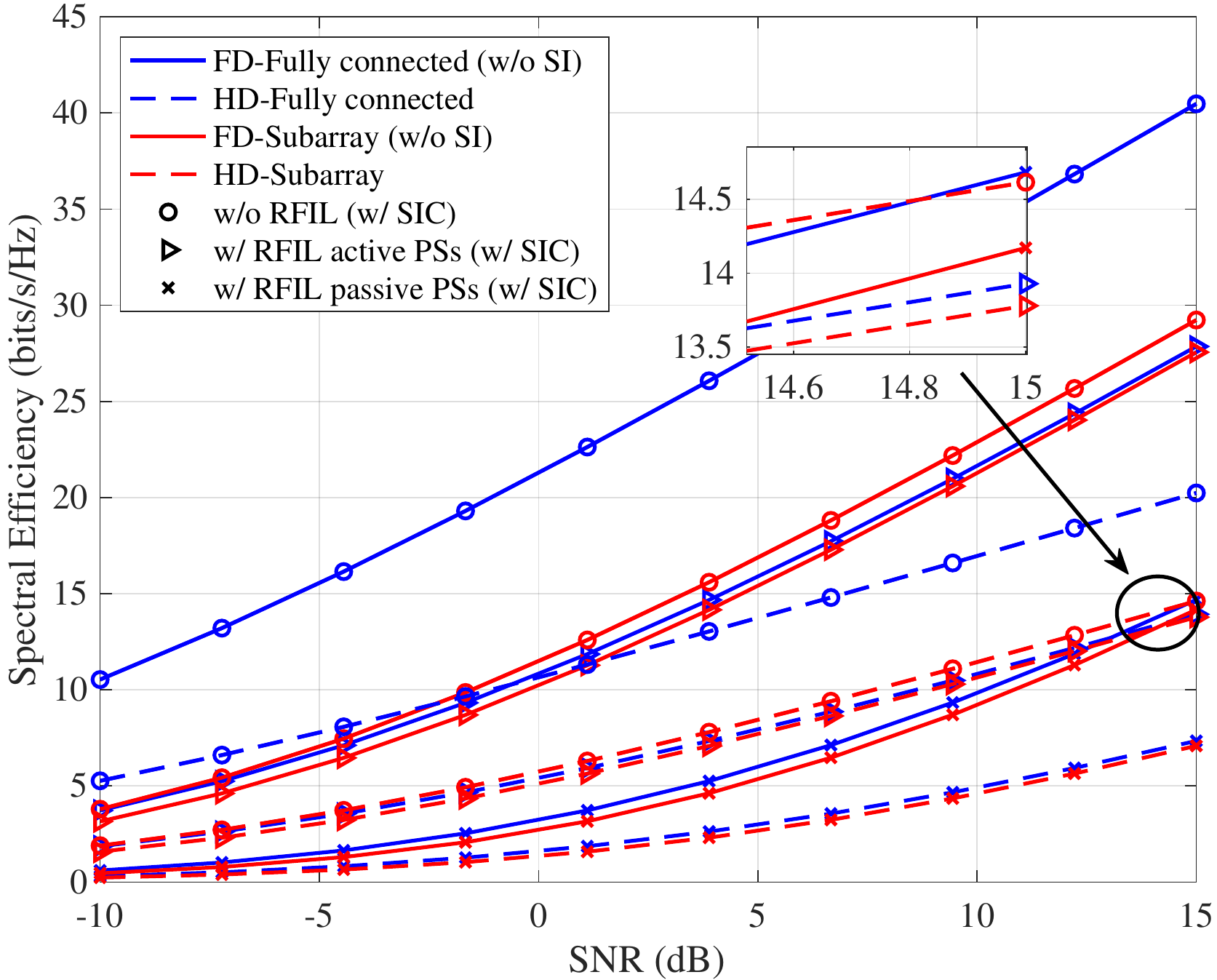}}
\caption{Comparison of the impact of RFIL on the SE of a 4-user mmWave-FD-IAB system with different hybrid precoding structures in terms of different kinds of PSs. The number of subarrays is equal to that of the user: a) backhaul link: 16 $\times$ 16 UPA, 4 (8) RF chains at Tx (Rx), 4 data streams; b) access link: 16 $\times$ 16 UPA and 4 RF chains at the Tx. Each user is equipped with 1 RF chain and 4 $\times$ 16 UPA and receives 1 data stream from Tx.}
\label{HI}
\end{figure}
\section*{Simulation Results}
In this section, simulations are presented to analyze the SE for our hybrid precoding design with the impact of the CEE and RFIL. The OFDM system has $K= 512$ subcarriers, where each channel realization has $D= 128$ delay taps. For a 4-subarray (user) hybrid precoding system, each subarray (user) has $4\times16$ UPA with 1 RF chain and 1 data stream. For successful digital cancellation, each subarray has two RF chains at the receiver of the IAB node. We assume that 80 dB SIC has been applied before the digital cancellation by the antenna and the analog cancellation \cite{Zhan2012}. We define signal-to-noise ratio (SNR)$\triangleq P_\mathrm{r}/(KU\sigma^2_\mathrm{n})$, where $P_\mathrm{r}=P_\mathrm{t}/\bar{PL}$ is the ratio between transmit power and average path loss according to the Friis’ law, and $\sigma^2_\mathrm{n}$ denotes the Gaussian noise power.

\subsection{Effect of RF Insertion Loss}
Figure~\ref{HI} shows the SE of both the backhaul and the access link with different hybrid precoding schemes by comparing FD and HD transmission in the presence of the RFIL in terms of different kinds of PSs. Both subfigures show a similar trend. Without considering the impact of the RFIL, the SE with FD transmission of the fully connected structure is much higher than that of the subarray structure, which has a difference of around 20 b/s/Hz and 12 b/s/Hz for the backhaul and the access links, respectively, at SNR $=15$ dB. For the HD scheme, this difference reduces to a half. However, in the presence of the RFIL, the SE obtained from the subarray structure is close to that given by the fully connected one, which means that our precoding scheme experiences less effect from the RFIL. Moreover, it can be seen that the use of active PSs can provide a higher SE than that with passive PSs, but with more power consumption \cite{8333733}. Specifically, for the backhaul link with ideal RF components, the SE of FD with SIC is close to the ideal one (i.e., with perfect SIC), which indicates successful SIC.
\begin{figure*}
\centering
\subfigure[]{
\includegraphics[width=\columnwidth]{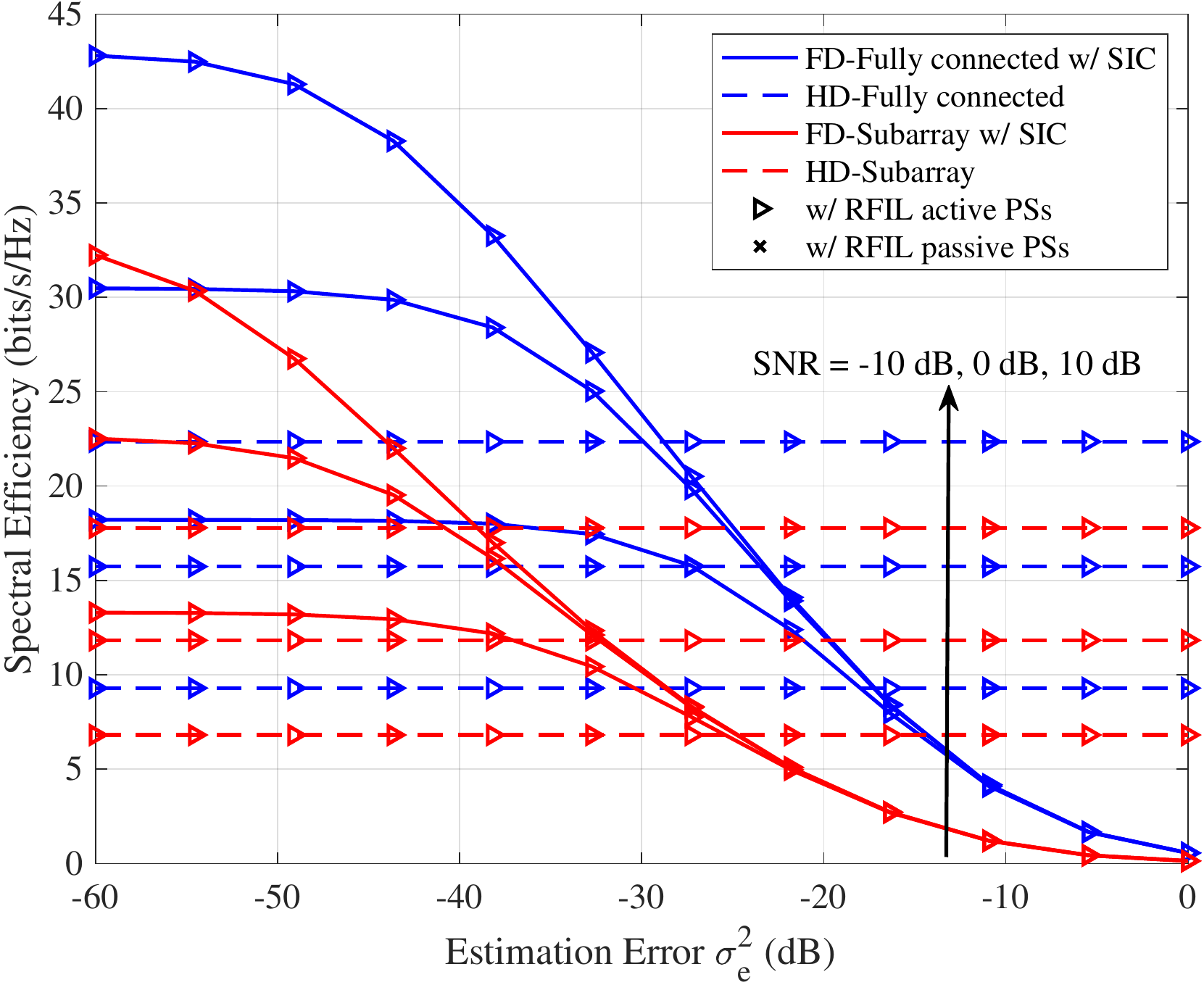}}
\subfigure[]{
\includegraphics[width=\columnwidth]{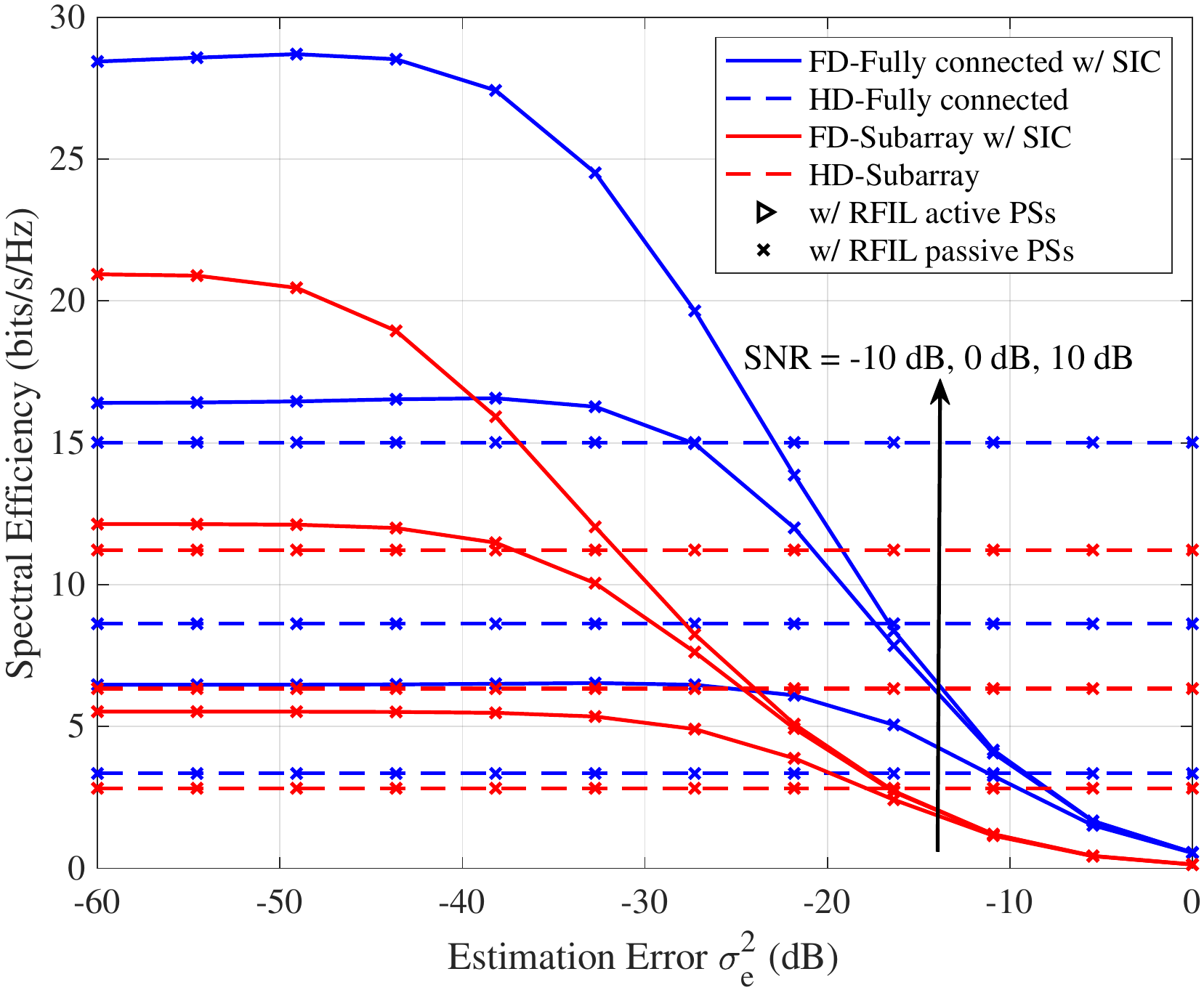}}
\caption{Comparison of the impact of CEE on the backhaul link SE in the presence of RFIL of a 4-user mmWave-FD-IAB system with different hybrid precoding structures in terms of different SNR values. Equipped with 16 $\times$ 16 UPA, 4 (8) RF chains at Tx (Rx), 4 data streams are transmitted. The number of subarrays is equal to that of the user: a) with active PSs; b) with passive PSs.}
\label{est}
\end{figure*}

\subsection{Effect of Channel Estimation Error}
We assume that only the RF effective SI channel is known with uncertainty. Therefore, only the backhaul link performance will be affected by the CEE. From Fig.~\ref{est}, it can be observed that irrespective of the selection of PSs, the higher SNR shifts the SE intersection of FD and HD to the left. At the right of the intersection, the FD scheme has less SE than the HD one due to the higher CEE. Moreover, compared to the fully connected structure, our subarray-based hybrid precoding scheme is more sensitive to the CEE. Therefore, more advanced techniques are needed to estimate the RF effective SI channel as accurately as possible. Further, interestingly, with passive PSs, the intersection points shift to the right, as compared to that for active PSs, implying the more tolerance of the system with passive PSs. It can be noted that although the fully connected structure shows better SE, the incurred hardware cost is much less for the subarray structure.

\subsection{Effect of RF Chains on Digital SIC}
\begin{figure}[t!]
\centerline{\includegraphics[width=\columnwidth]{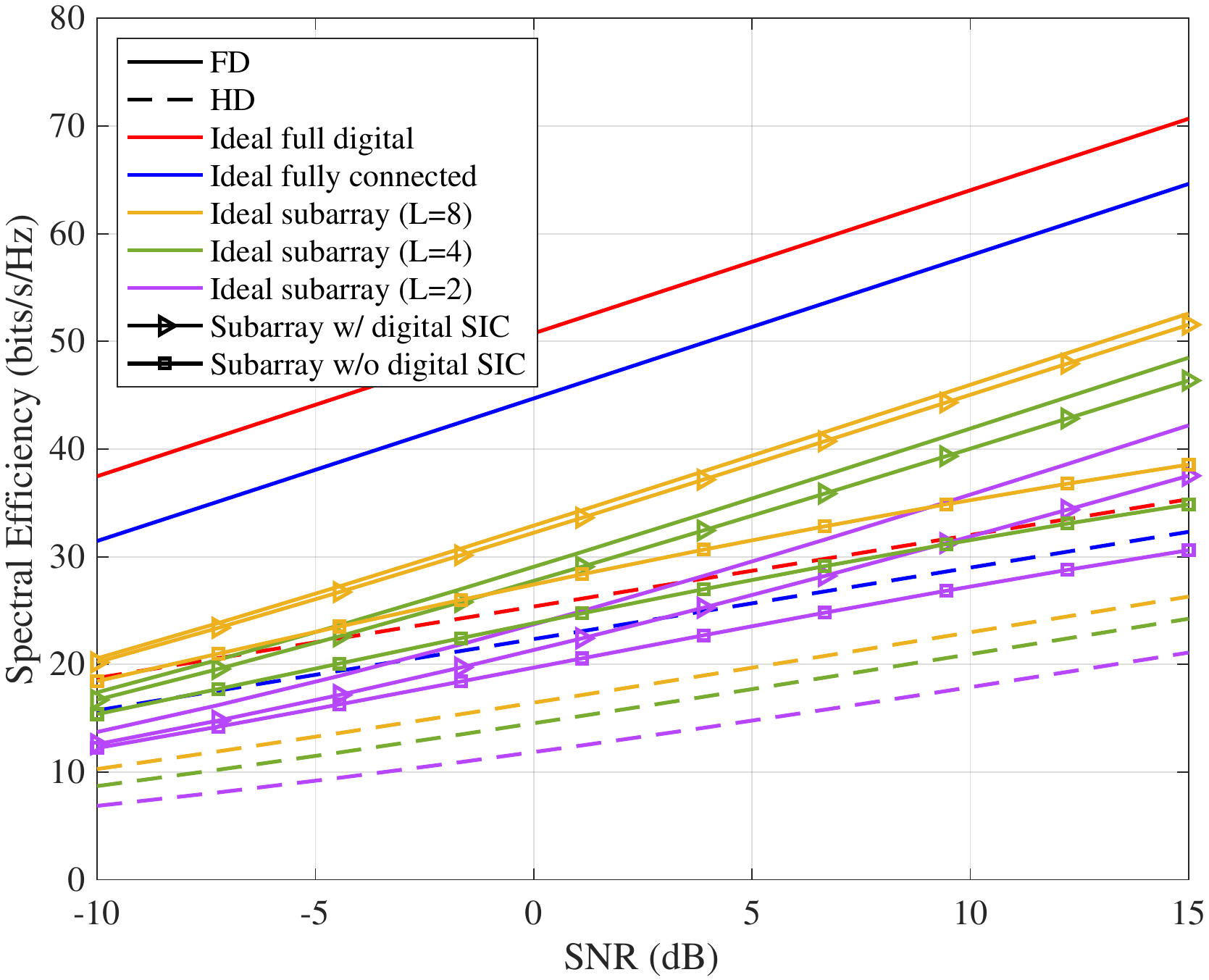}}
\caption{Digital SIC ability of a 4-user wideband mmWave-FD-IAB subarray network in terms of different numbers of RF chains $(L=2,4,8)$ on each Rx subarray. Equipped with 16 $\times$ 16 UPA on both sides, 4 RF chains at Tx and 4 data streams are transmitted.}
\label{backse}
\end{figure}

In Fig.~\ref{backse}, the digital SIC ability in terms of the SE of the backhaul link is plotted with different numbers of RF chains at the IAB node receiver. The fully connected hybrid precoding schemes are assumed to have 4 (8) RF chains at the transmitter (receiver). The ideal curves are plotted by assuming perfect SIC. It is evident that the ideal fully connected precoding provides close performance to the ideal full digital scheme, and leaves a gap with respect to the ideal subarray-based precoding scheme. Regarding the digital cancellation ability of the subarray structure, the more RF chains at the receiver of the IAB node, the more improvements in the SE can be seen and the smaller the SE difference with respect to the ideal subarray curves.At 15 dB SNR, with different numbers of RF chains at the receiver of the IAB node ($L = 2, 4, 8$ per subarray), the SE of deploying digital SIC is improved nearly 23, 33, and 34 percent, respectively, and the corresponding rate loss gets to around 4.7, 2.1, and 1 b/s/ Hz, respectively.

\section*{Conclusion}
In this article, we have presented the multiuser
mmWave-FD-IAB architecture according to the latest 3GPP standard for IAB networks. Wideband and FD operations have been investigated from the SE perspective. Further, the general mmWave channel model is described, followed by the characterization of the SI channel for mmWave FD operation, including the challenges in the SI channel estimation. Through a hardware cost-effective and computationally efficient subarray-based hybrid precoding scheme, with the objective of SE maximization in the IAB networks, MUI and RSI are mitigated at the IAB node transmitter and receiver using BB ZF and MMSE, respectively. The impact of the RFIL with active or passive PSs has been analyzed. To observe the effect of the imperfect RF effective CSI, the SE is plotted for different values of CEE in the presence of the RFIL and compared with the HD operation. Simulations have shown that if the CEE is inversely proportional to SNR, improvement of FD and HD can be observed. Moreover, the system with passive PSs can tolerate higher CEE than the system with active PSs.

Since the subarray hybrid precoding scheme is sensitive to CEE, adjustments need to be investigated for accurate RF effective SI channel estimation. Further, the equal power allocation assumption can be relaxed, and optimal power can be allocated to the effective channel. In practice, the PSs are not continuously controlled. Therefore, we will focus on quantization schemes with an efficient codebook design in the future. Moreover, an efficient antenna and RF cancellation are important to investigate to leverage the advantages of FD transmission.
\bibliographystyle{IEEEtran}
\bibliography{IEEEabrv,Ref}

\section*{Acknowledgment}
The work was supported in part by the research grant from Huawei Technologies
(Sweden) AB.

\section*{Biographies} 
\noindent \textsc{Junkai Zhang} [S'21] (jzhang15@ed.ac.uk) received his B.Eng. degree in communication engineering from Shenyang Ligong University, China, in 2018, and an M.Sc. degree in signal processing and communications, with Distinction, from The University of Edinburgh, United Kingdom, in 2019. He is currently with Institute for Digital Communications, The University of Edinburgh as a Ph.D. candidate. His research interests include 5G and beyond wireless networks, millimeter-wave communications, full duplex radio, and massive MIMO.
\\

\noindent \textsc{Navneet Garg} [S'15, M'19] (ngarg@ed.ac.uk) received his B.Tech. degree in electronics and communication engineering from the College of Science \& Engineering, Jhansi, India, in 2010, and his M.Tech. degree in digital communications from ABV-Indian Institute of Information Technology and Management, Gwalior, in 2012. He completed his Ph.D. degree in June 2018 from the Department of Electrical Engineering at the Indian Institute of Technology Kanpur, India. From July 2018 to January 2019, he visited The University of Edinburgh. From February 2019 to February 2020, he was employed as a research associate at Heriot-Watt University, Edinburgh. Presently, he is working as a research associate at The University of Edinburgh. His main research interests include interference alignment, edge caching, optimization, and machine learning.\\ 

\noindent \textsc{Mark Holm} [S'98, M'01] (mark.holm@huawei.com) received his B.Sc. (hons) in laser physics and optoelectronics from the University of Strathclyde in 1997, before studying for his Ph.D. in physics from the University of Strathclyde, which he received in 2001. He currently works as a technical lead and hardware system architect for Huawei Technologies (Sweden) AB with interests in microwave radio, phased array antennas, full duplex radio systems, and photonic radios. In the past, he was the microwave lead on AESA Radar systems, senior engineer responsible for GaAs pHemt modeling, and also a laser and package design engineer for SFP/XENPACK Fibre modules. He is published in the fields of laser design and GaAs device modeling.\\

\noindent \textsc{Tharmalingam Ratnarajah} [S'94, A'96, M'05, SM'05] (T. Ratnarajah@ed.ac.uk)  is currently with Institute for Digital Communications, The University of Edinburgh as a professor in digital communications and signal processing. He has supervised 15 Ph.D. students and 21 postdoctoral research fellows and raised more than US\$ 11 million+ of research funding. He was the Coordinator of the EU projects ADEL (\EUR 3.7 million) in the area of licensed shared access for 5G wireless networks, HARP (\EUR 4.6 million) in the area of highly distributed MIMO, the EU Future and Emerging Technologies projects HIATUS (\EUR 3.6 million) in the area of interference alignment, and CROWN (\EUR 3.4 million) in the area of cognitive radio networks. His research interests include signal processing and information theoretic aspects of 5G and beyond wireless networks, full duplex radio, mmWave communications, random matrix theory, statistical and array signal processing, and quantum information theory. He has published over 400 articles in these areas and holds four U.S. patents. He is a Fellow of the Higher Education Academy (FHEA). He was an Associate Editor of \textit{IEEE Transactions on Signal Processing} from 2015 to 2017 and Technical Co-Chair of the 17th IEEE International Workshop on Signal Processing Advances in Wireless Communications, Edinburgh, in 2016.
\end{document}